# Surface modes and chiral symmetry (Wilson Fermions in a box) *

Michael Creutz and Ivan Horvath

Physics Department, Brookhaven National Laboratory, PO Box 5000, Upton, NY 11973-5000, USA
creutz@bnl.gov, horvath@wind.phy.bnl.gov

We give a Hamiltonian discussion of surface states in an extra dimension as a basis for chiral fermions in lattice models. Such modes appear with the Wilson fermion action when the hopping parameter exceeds a critical value. The association of such states with the closing and reopening of a band gap was noted by Shockley in 1939.

Chiral symmetry in the lattice framework is a longstanding and classic problem. For a review see Petcher's contribution to last year's meeting [1]. Here we present our efforts to understand recent approaches based on fermionic states bound to defects in a higher dimensional space.

As is well known, simple attempts to place fermionic fields on a lattice tend to yield species beyond those initially intended. One traditional scheme for removing these infamous "doublers" involves adding terms which naively vanish in the continuum limit but give a large energy to the extra states. Unfortunately, this explicitly violates chiral symmetry; so, the usual approach is to tune the parameters to make the pion mass small and hope that the predictions of current algebra will be recovered in the continuum limit. Given the historical importance of chiral symmetry to our understanding of particle physics, this artificial prescription is not particularly satisfying.

The last year has seen considerable activity on using an infinite number of regulator fields to solve this problem [2, 3]. One particularly elegant realization of this approach involves the use of Shockley surface states [4].

We first review the standard Wilson fermion approach. Working in one space dimension for

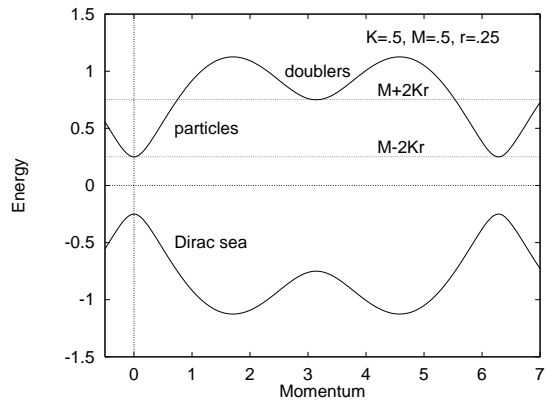

Figure 1. The spectrum of free Wilson fermions.

simplicity, consider the Hamiltonian

$$H = K \sum i(a_j^\dagger a_{j+1} - b_j^\dagger b_{j+1}) + \text{h.c.}$$
$$+ M \sum a_j^\dagger b_j + \text{h.c.} \qquad (1)$$
$$- Kr \sum (a_j^\dagger b_{j+1} + b_j^\dagger a_{j+1}) + \text{h.c.}$$

Here $a$ and $b$ represent fermion annihilation operators on a chain of sites labeled by the index $j$. The quantity $K$ is the "hopping parameter." In momentum space the single-fermion eigenstates have energies $E$ satisfying

$$E^2 = 4K^2 \sin^2(q) + (M - 2Kr\cos(q))^2 \qquad (2)$$

where $0 \le q < 2\pi$. This spectrum is sketched in Fig. 1. The physical vacuum has the negative energy states filled. The "Wilson term" proportional to $r$ makes the "doublers" at $q \sim \pi$ heavier than the states at $q \sim 0$.

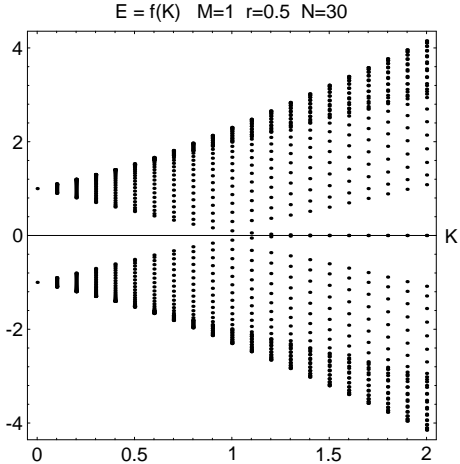

Figure 2. Energy levels versus hopping parameter $K$ for Wilson fermions on a 30 site lattice.

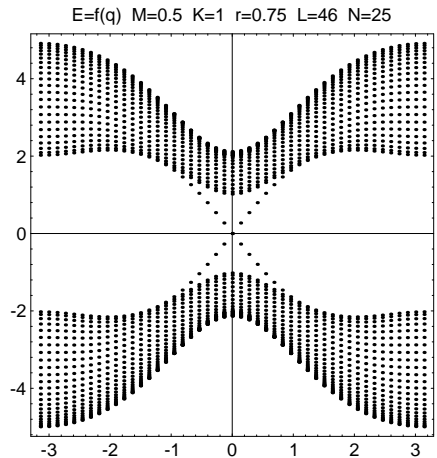

Figure 3. The energy spectrum as a function of the physical momentum on a lattice with 46 space sites and 25 sites in the extra dimension. Note the crossing surface modes at low energy.

At the critical value $2Kr = M$ the gap closes and one species of fermion becomes massless. While the symmetry is not exact when the lattice spacing is finite, this represents a candidate for a chirally symmetric theory in the continuum limit. Beforehand, as discussed in Ref. [5], chiral symmetry does not provide an order parameter. A further difficulty is that gauge interactions will renormalize the parameters. To obtain massless pions one must finely tune $K$ to $K_{crit}$, an apriori unknown function of the gauge coupling.

The free Wilson theory has interesting properties in the supercritical case, where $K > K_{crit} = M/2r$. As $K$ increases through the critical value, the gap in the spectrum first closes and then reopens. If we work in a large box with open walls the final spectrum consists of a particle band with $E > 0$, an antiparticle band at negative energy which represents the filled Dirac sea, and finally two surface states near $E = 0$ bound to the box walls. This behavior is plotted in Fig. 2. This figure has a close similarity to Fig. 2 of Ref. [4].

A more general result is that there will exist similar states bound to any interface separating a region with $K > K_{crit}$ from a region with $K < K_{crit}$. In Ref. [2], Kaplan uses $M = 2Kr + m\epsilon(x)$. We adopt the simpler approach of Shamir [6] and take $K = 0$ on one side, giving modes on an open surface. The energy of these states goes to zero as the box becomes infinite in length.

For a theory of chiral symmetry, we now turn the picture on its side and regard the above hopping as occurring in an extra "fifth" dimension. We live on a four dimensional "interface" with physically observed particles being surface modes. Zero mass fermions become a natural consequence of the vanishing energy of the surface states. With a finite but large box, opposite walls give rise opposite helicity states. Anomalies arise quite naturally as tunnelling between the walls. A continuum version of this phenomenon was presented in Ref. [7].

When physical dimensions are added, two things happen. First, physical momentum moves the surface modes from $E = 0$. In one physical space dimension they go to $E = \pm\sin(q_x)$, with leftmovers and right-movers living on opposite walls in the extra dimension. This is illustrated in Figs. 3 and 4.

Second, because of the Wilson term in the space directions, $K_{crit}$ depends on the physical momentum. The condition for surface modes to exist then depends on the spatial momentum. If they disappear before reaching momentum $\pi$, as

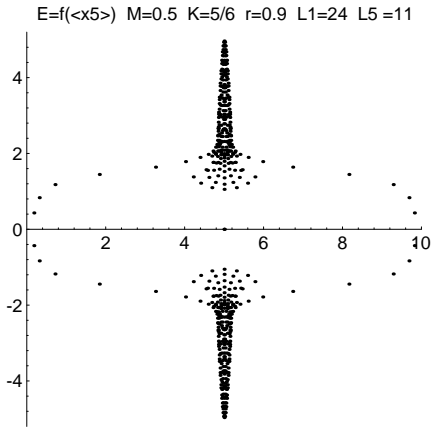

Figure 4. The energy spectrum as a function of the average position in the extra dimension on a $N_1 = 24$ by $N_5 = 11$ lattice. Note how the low energy modes lie near the ends of the lattice.

seen in Fig. 3, doublers are eliminated. For one space dimension we have

$$K_{crit} = M/2r - K \cos(q_x) \qquad (3)$$

and the doubler at $q_x = \pi$ is absorbed in continuum of the fifth dimension if

$$M/2r - K < K < M/2r + K. \qquad (4)$$

This explains the pictures in Ref. [8].

Ultimately interest lies in coupling the fermions to gauge fields. To avoid extraneous fields from the extra dimension, it is simplest to have gauge fields in physical dimensions only. We also put the same field with a given physical position at all $x_5$; thus, as in Ref. [3], $x_5$ represents a "flavor" coordinate. With this prescription, the gauge field has equal couplings to the opposite chirality states on the opposing walls and effectively we have a Dirac fermion. With two flavors the scheme should give rise to massless pions with no tuning or doubling. Quarks can acquire a mass via an explicit coupling between the opposite walls.

An extension of these ideas to chiral gauge theories remains open. Weak interactions violate parity, and we would like a non-perturbative formulation. In practice this is irrelevant since the electroweak couplings are small and perturbation theory works quite well. On the other hand, the lattice is the cleanest non-perturbative regulator known and we would like to understand all interactions in a more fundamental way. Any valid formulation must cancel anomalies in gauged currents. For the surface mode picture to work for the standard model, baryon number non-conservation through instantons [9] should arise from tunnelling through the fifth dimension.

We now discuss a toy model with mirror fermions. Consider two species $\psi_1$ and $\psi_2$ in the surface mode picture. Flip appropriate signs in the Hamiltonian so that they have opposite chirality on a given wall. Since we want to eventually couple only one handed neutrinos to the vector bosons, consider gauging $\psi_1$ but not $\psi_2$. We can now generate masses as in the standard model by coupling $\psi_1$ and $\psi_2$ through a Higgs field.

The new feature is that now the coupling to the Higgs can depend on the extra coordinate $x_5$. In particular, let it be small on one wall and large on the other. The fermions are then light on one wall and heavy on the other. This model is equivalent to the picture in Golterman, Jansen, Petcher and Vink [10], where the gauge fields are turned off in the middle of the slab, and heavy fermions appear bound to this discontinuity. The resulting model has a light chiral fermion and a heavy mirror fermion on the opposite wall. As in other mirror fermion models [11], triviality arguments suggest that the heavy particle cannot become much heavier than the vector mesons, i.e. the $W$.

A speculative interpretation would be to call the light fermion a lepton and the heavy fermion an antibaryon. Then one would have baryon decay occurring through tunnelling in the fifth dimension, while $B - L$ is still naturally conserved. Any simple extension of this idea to a realistic model should unify leptons and baryons. Perhaps one can use $SO(10)$ [12].

Now we turn to a technical discussion of the general domain wall solution. Consider one particle states for the Hamiltonian of Eq. (1)

$$|\psi\rangle = \sum_j (\psi_+(j) a_j^\dagger + \psi_-(j) b_j^\dagger) |0\rangle. \qquad (5)$$

In the gap with $|E| < |2Kr - M|$, we look for

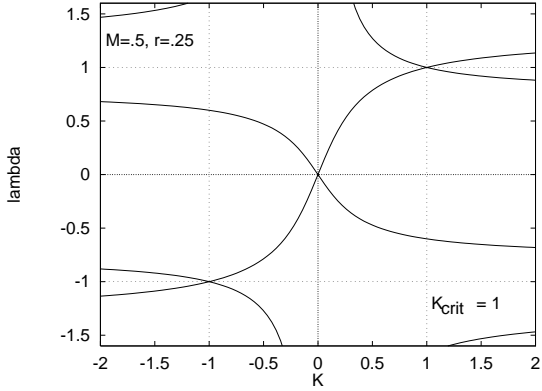

Figure 5. The translation eigenvalues at zero energy as a function of the hopping parameter. Note the crossing at the critical hopping $K = 1$. Here $r = .25$ and $M = .5$.

exponential solutions

$$\psi(j) \sim \lambda^j \psi(0). \tag{6}$$

Given an energy $E$, there are four possible values for $\lambda$ satisfying

$$E^2 = -K^2(\lambda - 1/\lambda)^2 + (M - Kr(\lambda + 1/\lambda))^2 \tag{7}$$

For any solution $\lambda$, $1/\lambda$ also is. Thus two roots exponentially decrease, and two increase. If we specialize to $E = 0$ the equations simplify

$$\begin{aligned} -iK(\lambda - 1/\lambda)\psi_+ &= (M - 2Kr(\lambda + 1/\lambda))\psi_- \\ iK(\lambda - 1/\lambda)\psi_- &= (M - 2Kr(\lambda + 1/\lambda))\psi_+ \end{aligned} \tag{8}$$

This immediately implies that $\psi_- = \pm i\psi_+$.

Fig. 5 shows the behaviour of the four eigenvalues as a function of the hopping parameter $K$. A crucial eigenvalue crossing occurs at $K = K_{crit}$. For $|K| > |\frac{M}{2r}|$, the two exponentially decreasing solutions have the same phase relation $\psi_- = -i\psi_+$, while the two increasing solutions have $\psi_- = i\psi_+$. When $|K| < |\frac{M}{2r}|$ the solutions pair oppositely; for a given phase relation there is both an increasing and a decreasing solution.

At a wall separating subcritical and supercritical hopping, one can match a linear combination of the two supercritical solutions onto the subcritical one with the same phase between $\psi_-$ and $\psi_+$. With the appropriate choice for this phase, the resulting solution will be normalizable.

For the case of an open wall, consider a semi-infinite box with supercritical $K$ for $j \geq 1$ and $K$ vanishing for $j \leq 0$. The zero mode solution for positive $j$ is then $\psi_- = -i\psi_+$ and

$$\psi_+(j) \sim (\lambda_1^j - \lambda_2^j) \tag{9}$$

where $\lambda_1$ and $\lambda_2$ represent the two decreasing eigenvalues. This combination automatically satisfies the boundary condition of vanishing at $j = 0$. In a finite box an exponentially suppressed mixing of the surface modes will generally give the states a small nonvanishing energy.

**Acknowledgement:** We thank J. Davenport for pointing out that these surface modes are Shockley states.